\documentclass[rnote]{gaa}
\usepackage{graphicx}
\usepackage{longtable}
\usepackage{supertabular}

\def\arcmin{\tt '}
\def\arcsec{\tt ''}
\def\simgt{\ {\raise-.5ex\hbox{$\buildrel>\over\sim$}}\ }
\def\simlt{\ {\raise-.5ex\hbox{$\buildrel<\over\sim$}}\ }

\def\cd{d$^{-1}$\,}

\begin{document}

\title{New $\beta$~Cephei stars in the young open cluster NGC~637}
\authorrunning{G. Handler \& S. Meingast}
\titlerunning{New $\beta$~Cephei stars in NGC~637}
   \author{G. Handler,$^{1,2}$ S. Meingast$^2$}
   \offprints{G. Handler}

   \institute{
$^1$ Copernicus Astronomical Center, Bartycka 18, 00-716 Warsaw, Poland (gerald@camk.edu.pl)\\
$^2$ Institut f\"ur Astronomie, Universit\"at Wien, 
T\"urkenschanzstra\ss e 17, 1180 Wien, Austria}

\date{Received March 14, 2011; Accepted May 16, 2011}

\abstract
{Studying stellar pulsations in open clusters offers the possibility to 
perform ensemble asteroseismology. The reasonable assumption that the 
cluster members have the same age, distance, and overall metallicity 
aids in the seismic modelling process and tightly constrains it. 
Therefore it is important to identify open clusters with many 
pulsators.}
{New pulsating stars of the $\beta$~Cephei type were searched for 
among the members of the open cluster NGC~637.}
{Thirty-one hours of time resolved $V$ filter CCD photometry were obtained.}
{The measurements confirmed two previously known variables, and revealed 
three new $\beta$~Cephei stars plus one more candidate. All four 
pulsators have amplitudes high enough for easy mode identification and are 
multiperiodic.}
{With four certain pulsating members, NGC 637 is now among the six open 
clusters richest in $\beta$~Cephei stars. It is thus an excellent target 
for ensemble asteroseismology, and to tackle the question what separates 
pulsating from apparently constant stars in the $\beta$~Cephei domain.}

\keywords{Stars: early-type - open 
clusters and associations: general  - Techniques: photometric - Stars: 
oscillations - Asteroseismology}

\maketitle

\section{Introduction}

Asteroseismology is the study of the interiors of pulsating stars by 
using their modes of oscillation as seismic waves. The mode frequencies 
depend on the physical conditions in the regions where the oscillations 
propagate. Measuring these frequencies therefore provides information 
about the deep inner structure of stars, impossible in any other way. An 
overview of asteroseismology, its methods and results can be found in 
the monograph by Aerts, Christensen-Dalsgaard \& Kurtz (\cite{ACK10}).

To make asteroseismic investigations possible, one needs to know the 
surface and interior geometries of the observed oscillation modes.
This poses the problem of mode identification (e.g., see Handler 
\cite{H08}, Telting \cite{T08}). In addition, knowledge of the effective 
temperatures and luminosities of the target stars is important. However, 
absolute magnitudes of field stars can often only be determined with 
unsatisfactory accuracy.

The study of pulsating stars in clusters and associations alleviates 
this problem. Not only can the distances to stellar aggregates be 
determined accurately, but as stars belonging to open clusters have 
originated from the same interstellar cloud, they can be assumed to have 
the same age and metallicity. Consequently, the asteroseismic modelling 
process of pulsating stars in clusters is tightly constrained.

The potential of asteroseismology has been recognized for the $\beta$ 
Cephei stars (e.g., Pamyatnykh, Handler, \& Dziembowski \cite{PHD04}), 
that are pulsating main sequence variables with early B spectral types 
and typical periods of several hours (Stankov \& Handler \cite{SH05}, 
Pigulski \& Pojma{\'n}ski \cite{PP08}). These young massive stars 
naturally occur in open clusters and stellar associations, and extensive 
observing campaigns have been organized (e.g., Saesen et al. 
\cite{SP10}).

Searches for $\beta$ Cephei stars in open clusters have increased the 
total number of such objects to about 50 out of a total of some 200 
galactic $\beta$ Cephei stars known (Pigulski \cite{PA06}, Pigulski \& 
Pojma{\'n}ski \cite{PP08}). The relative number of pulsators to 
nonpulsators is considerably lower in northern hemisphere open clusters, 
which has been attributed to a metallicity gradient in our Galaxy 
(Pigulski \cite{P04}).

NGC 637 is a young (age $10\pm5$ Myr, Yadav et al. \cite{Y08}) open 
cluster whose seven brightest stars are of 10$^{\rm th}$ magnitude. 
Apart from some radial velocity measurements (Liu, Janes, \& Bania 
\cite{LJB91} and earlier papers) this cluster has only been 
photometrically studied. In particular, no membership information is 
available in the literature.

NGC~637 has been examined for stellar variability. Two of the seven 
brightest stars turned out to variable (Pietrukowicz et al.\ 
\cite{P06}). One has been classified as a $\beta$~Cephei star, and the 
other was suggested to be an ellipsoidal variable. New standard 
$uvby\beta$ photometry (Handler \cite{H11}) puts all seven bright stars 
into the $\beta$~Cephei instability strip (Fig.\ 1). Therefore, and to 
evaluate the prospects of NGC~637 as a target for ensemble 
asteroseismology, it was decided to re-examine the cluster for pulsating 
stars.

\begin{figure}
\includegraphics[width=88mm,viewport=00 00 265 260]{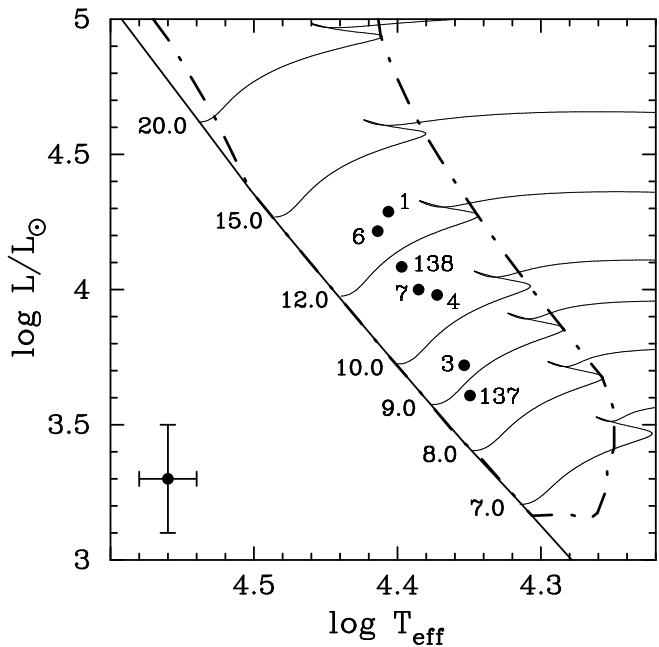}
\caption{The positions of the seven brightest stars in NGC~637, as 
derived in Sect.\ 4, and labelled with their WEBDA ({\tt 
http://www.univie.ac.at/webda/}) numbers, in a $\log T_{\mathrm{eff}} - 
\log L$ diagram. Some model evolutionary tracks computed with the 
Warsaw-New Jersey code (e.g., see Pamyatnykh et al. \cite{P98}) are 
shown for comparison, marked with the corresponding masses. The 
slanted full line is the zero age main sequence. The dashed-dotted 
line is the theoretical $\beta$~Cephei star instability strip (Zdravkov 
\& Pamyatnykh \cite{ZP08}). Error bars of the individual positions 
of the stars are given in the lower left corner.}
\end{figure}

\section{Observations and data reduction}

Time series photometric measurements were carried out with the 0.8-m 
``vienna little telescope" (vlt) at the Institute of Astronomy of the 
University of Vienna. The telescope is located at an altitude of 241\,m, 
some 5\,km away from the city centre. It was used in combination with an 
SBIG STL-6303E CCD camera, with a focal reducer and a 3K $\times$ 2K 
front-illuminated Kodak Enhanced KAF-6303E chip. The pixel size of $9 
\times 9$ microns translates to $0.41\arcsec$ on the sky and gives a 
total field of view of $20.8\arcmin \times$\,$13.9\arcmin$. The measured 
value for the gain is $2.32 \pm 0.02$\,e$^-$/ADU in $2\times2$ binning, the 
read noise was determined with 11.2\,e$^-$ and the dark current with 0.07 
e$^-$/pix/s at $0\degr$\,C; the two latter values are better than the 
official camera specifications.

Given the typical seeing in Vienna (between $1.4\arcsec -$\,$6\arcsec$ 
FWHM, with a median of $2.8\arcsec$ during the present measurements), 
the chip was binned $2\times2$, resulting in a readout time of 8\,s. Due 
to technical problems with the filter wheel, only the $V$ filter was 
used. Integration times of 10 or 15\,s were chosen, depending on seeing. 
NGC~637 was observed on nine nights from December 2010 to March 2011 in 
which a total of 4776 usable science frames was collected.

The journal of the observations is given in Table~1. Due to constraints 
from other observing programs, the camera was rotated twice during the 
project. Care was taken that after the second rotation the observed 
field matched the original orientation as closely as possible. The time 
base of the total data set is 61.2\,d.

\begin{table}
\begin{center}
\caption{Journal of the observations}
\begin{tabular}{ccc}
\hline
\multicolumn{2}{c}{Run start} & Length \\
Civil Date & UT & hr \\
\hline
30/12/2010 & 16:19:38 & 0.89\\
06/02/2011 & 16:51:19 & 4.99\\
\hline
07/02/2011 & 17:13:29 & 4.34\\
\hline
23/02/2011 & 17:19:55 & 4.25\\
24/02/2011 & 17:09:03 & 1.13\\
25/02/2011 & 17:11:58 & 4.37\\
26/02/2011 & 17:59:19 & 3.02\\
28/02/2011 & 17:16:56 & 4.34\\
01/03/2011 & 17:17:47 & 4.01\\
\hline 
Total & & 31.34 \\
\hline
\end{tabular}
\tablefoot{Times of CCD rotation are identified with horizontal lines.} 
\end{center}
\end{table}

The CCD frames were reduced with standard IRAF tasks, comprising 
corrections for bias level, dark current and flat field. Flat field 
frames were obtained by exposing the CCD to the evening twilight sky. 
Heliocentric Julian Dates were computed for the middle of each exposure.
Photometry of the reduced frames was carried out using the MOMF 
(Multi--Object Multi--Frame, Kjeldsen \& Frandsen \cite{KF92}) package. 
MOMF applies combined Point--Spread Function/Aperture photometry 
relative to an optimal sample of comparison stars, ensuring 
highest-quality differential light curves of the targets.

All sets of frames, grouped by field orientation,
were searched for variables as a first step. Then, the optimal 
apertures, in terms of minimizing the scatter in the resulting light 
curves, were determined for all targets of interest from the most 
extensive subset of data (Feb 23 $-$ Mar 1). These apertures were then 
used for all data subsets, and differential magnitudes were computed for 
all stars. The light curves of the data subsets were combined and the 
relative zeropoints were adjusted between them as the comparison star 
sample varied with field orientation, but not between individual nights. 
The resulting final light curves were cleaned for statistically 
significant outliers and first searched for variability up to the 
Nyquist frequency. No evidence for periodic signals with periods between 
47\,s and 90\,min was found. Consequently, the data were merged into 
3-minute bins and subjected to variability analysis.

\section{Analysis}

Unless otherwise noted, frequency analyses were performed with the 
program package {\tt Period04} (Lenz \& Breger \cite{LB05}). This 
software uses single frequency Fourier and multifrequency nonlinear 
least squares fitting algorithms. Amplitude spectra were computed and 
the frequencies of the intrinsic and statistically significant peaks in 
the Fourier spectra were determined. Multifrequency fits were calculated 
with all detected signals, and the corresponding frequencies, amplitudes 
and phases were optimized. The resulting fit was subtracted from the 
data before residual amplitude spectra were computed. These were 
examined in the same way until no significant further periodicity could 
be detected. In what follows, this procedure is called prewhitening.

The frequency solutions derived from the new light curves are given in 
the consequent subsections. On suggestion by the referee, formal 
error estimates of the individual parameters following Montgomery \& 
O'Donoghue (\cite{MO99}) are given. However, they will be systematically 
overoptimistic: our frequency analyses are affected by aliasing and 
more signals may be present in the data. 

\subsection{Previously known variable cluster members}

Pietrukowicz et al. (\cite{P06}) discovered the variability of 
NGC~637~4\footnote{All stars are identified with their WEBDA 
designations, whenever available.} and identified it as a $\beta$~Cephei 
pulsator. The light curves of this star from the present study are shown 
in the left part of Fig.\ 2. The light range varies from 0.1 to 0.2 mag 
in different nights, suggesting beating of several pulsation modes. This 
is confirmed by the Fourier analysis of the light curves (Fig.\ 2, right 
half). The corresponding frequency solution, leaving behind an rms 
residual of 3.4 mmag per single data point, is listed in Table 2 (top). The 
two strongest oscillations and the harmonic were previously found by 
Pietrukowicz et al. (\cite{P06}), consistent with the present results. 
The third oscillation frequency is new.

\begin{figure*}
\includegraphics[width=134mm,viewport=-55 00 445 260]{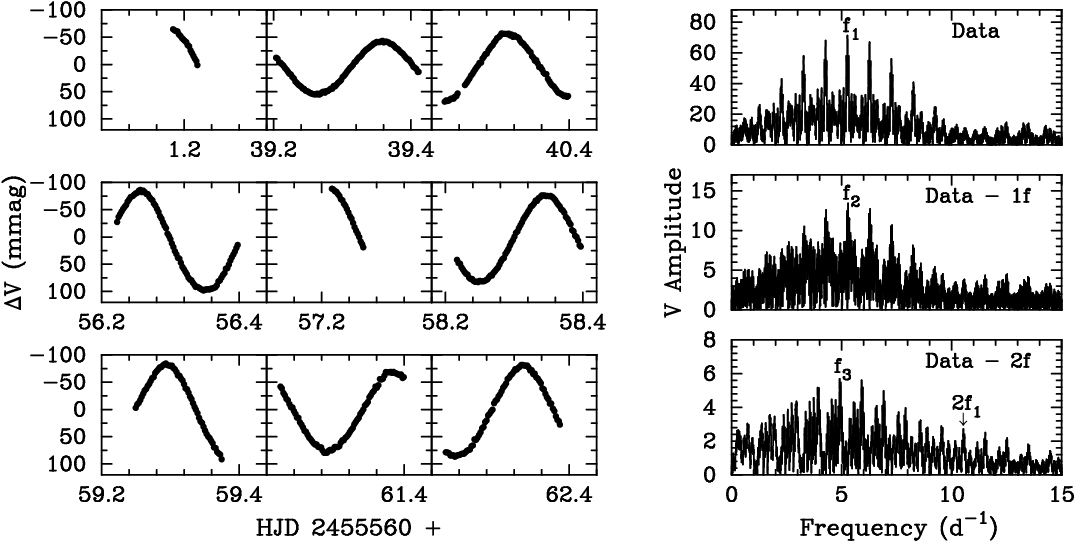}
\caption{Left: light curves of the known $\beta$~Cephei star 
NGC~637~4. Right: amplitude spectra of the data with successive 
prewhitening.}
\end{figure*}

NGC 637 3 was also recognized as a variable star with a main period of 
0.7432~d (Pietrukowicz et al. \cite{P06}). These authors preferred an 
explanation as an ellipsoidal variable (with an orbital period of 
$2\times0.7432$~d) for the star over that as a slowly pulsating B star 
because of its high luminosity.

Fourier analysis of the present data set reveals the same dominant 
period. However, pushing it further leads to difficulties because the 
subharmonic of this frequency (0.673 \cd) cannot be distinguished from 
its first harmonic (2.691 \cd) due to daily aliasing. Even worse, the $-2$ 
\cd alias of the strongest signal is not fully resolved from the 
possible subharmonic. Residualgram analysis (Martinez \& Koen 
\cite{MK94}), consisting of fitting a sine wave with $M$ harmonics to 
the measured time series and evaluating the residuals at each trial 
frequency, provides the solution (Fig.\ 3, upper panel). Fitting a 
signal with one harmonic to the data clearly shows that the base 
frequency is 0.673 \cd. The resulting frequency solution is given in 
Table~2 (bottom); no other signal is significantly present in the light 
curves. The rms of the residual time series is 2.2 mmag per point.

The lower panel of Fig.\ 3 shows a phase diagram of the light curve with 
the base frequency. Unfortunately, there is only about 56\% phase 
coverage. However, a fit to this phase diagram (cf.\ Fig.\ 3) suggests a 
light curve resembling an ellipsoidal variable, with equal light maxima 
and unequal minima. Although we cannot rule out that the 
variability is induced by rotation, we prefer the interpretation of 
ellipsoidal light variations for NGC~637~3.

\begin{figure}
\includegraphics[width=88mm,viewport=0 00 270 283]{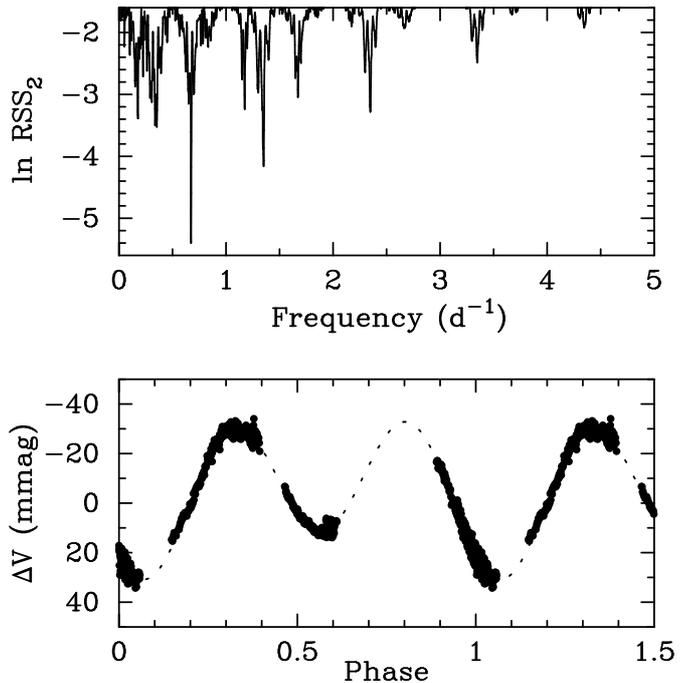}
\caption{Upper panel: residualgram analysis of the light curves of the 
previously known variable NGC~637~3. Lower panel: phase diagram of these 
data with the best fit to it (dotted line).}
\end{figure}

\begin{table}
\begin{center}
\caption{Frequency solutions for the known variable cluster members.}
\begin{tabular}{llccc} 
\hline
\# & ID & Frequency & V Amplitude & $S/N$ \\ 
 & & (\cd) & (mmag) & \\
\hline
4 & $f_1$ & $5.26350 \pm 0.00002$  & $65.2 \pm 0.2$ & 64.6 \\
 & $f_2$ & $5.28770 \pm 0.00007$  & $22.5 \pm 0.2$ & 22.6 \\
 & $f_3$ & $4.9073  \pm 0.0002$ &   $7.0  \pm 0.2$ & 6.7 \\
 & $2f_1$ & $10.52699 \pm 0.00004$ & $2.3 \pm 0.2$ & 4.1 \\
\hline
3 & $f_1$ & $0.67266 \pm 0.00008$ & $9.9 \pm 0.1$ & 13.5 \\
 & $2f_1$ & $1.34532 \pm 0.00004$ & $26.2 \pm 0.1$ & 36.5 \\
\hline
\end{tabular} 
\tablefoot{\# is the WEBDA identification of the stars. The error estimates are formal and should be used with caution.}
\end{center}
\end{table}

\subsection{New $\beta$ Cephei stars}

The brightest member of NGC~637, WEBDA 1, is a new $\beta$ Cephei star. 
Its light curves (Fig.\ 4, left-hand side) can reach light ranges of 
about 0.045 mag in $V$. Multiperiodicity is evident from the light 
variations and from a Fourier analysis with prewhitening (Fig.\ 4, 
right-hand side). The frequency analysis for these data is summarized in 
Table~3 (top). There are aliasing uncertainties, and the residual amplitude 
spectrum suggests the presence of more, or unresolved, frequencies in 
the light variations of the star.

\begin{figure*}
\includegraphics[width=134mm,viewport=-55 00 445 175]{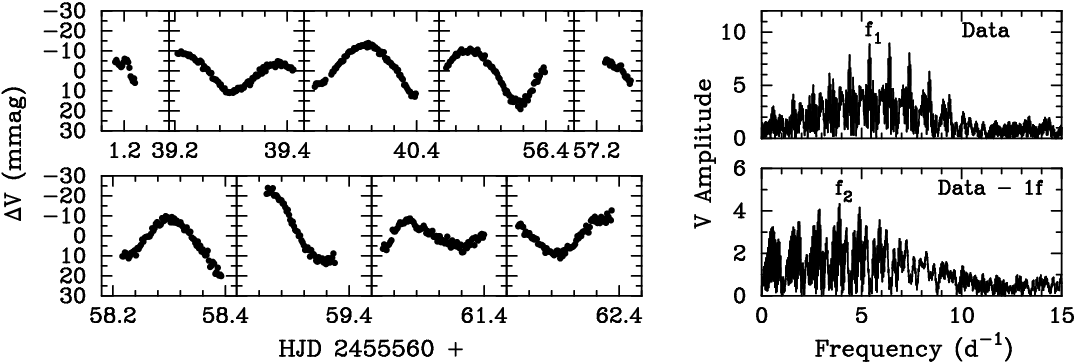}
\caption{Left: light curves of the new $\beta$~Cephei star NGC~637~1. 
Right: amplitude spectra of these data with successive prewhitening.}
\end{figure*}

Most of the previous comments apply to the variability of NGC~637~7 as 
well. The star is also multiperiodic with slightly smaller light range 
(up to 0.041 mag, Fig.\ 5). However, although the aliasing is as severe 
as for the previous star, it was easier to find the most probable 
intrinsic frequencies (Table~3, middle) and the prewhitening residuals are lower 
(2.8 vs.\ 3.7 mmag rms per point), suggesting that this star has a 
simpler oscillation behaviour than NGC~637~1.

\begin{figure*}
\includegraphics[width=134mm,viewport=-55 00 445 175]{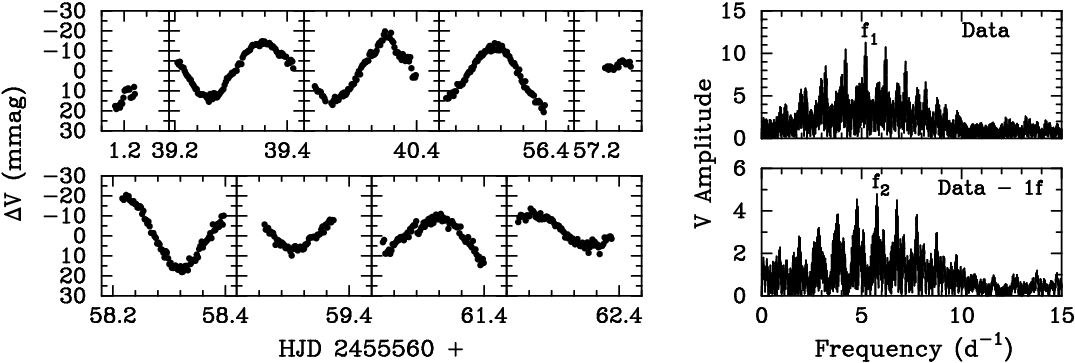}
\caption{Left: light curves of the new $\beta$~Cephei star NGC~637~7. 
Right: amplitude spectra of these data with successive prewhitening.}
\end{figure*}

The variability of NGC~637~138 is similar to the two stars discussed 
above. The V light range can reach 0.045 mag (Fig.\ 6). Results from the 
frequency analysis are listed in Table~3 (bottom); the rms residual 
scatter is 3.4 mmag per data point. A re-analysis of the time 
series photometry by Pietrukowicz et al.\ (\cite{P06}) shows the 
pulsations of these three stars to be marginally present, but these data 
alone were unfortunately insufficient for confident detections.

\begin{figure*}
\includegraphics[width=134mm,viewport=-55 00 445 175]{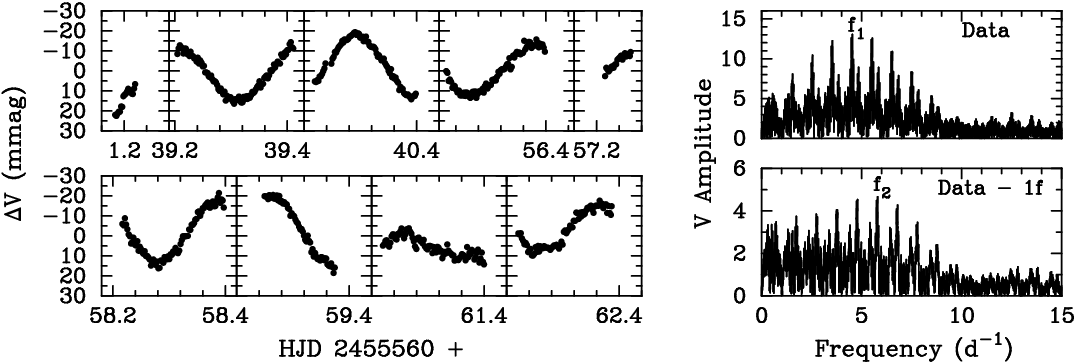}
\caption{Left: light curves of the new $\beta$~Cephei star 
NGC~637~138. Right: amplitude spectra of these data with successive 
prewhitening.}
\end{figure*}

\begin{table}
\begin{center}
\caption{Frequency solutions for the new $\beta$ Cephei stars in NGC 637.}
\begin{tabular}{llccc} 
\hline
\# & ID & Frequency & V Amplitude & $S/N$ \\ 
 & & (\cd) & (mmag) & \\
\hline
1 & $f_1$ & $5.3899 \pm 0.0002$ & $9.3 \pm 0.2$ & 8.0 \\
 & $f_2$ & $3.8962 \pm 0.0004$ & $4.5 \pm 0.2$ & 4.0 \\
\hline
7 & $f_1$ & $5.1982 \pm 0.0001$ & $10.7 \pm 0.2$ & 14.1 \\
 & $f_2$ & $5.7634 \pm 0.0003$ & $4.9 \pm 0.2$  & 7.0 \\
\hline
138 & $f_1$ & $4.5364 \pm 0.0002$ & $11.5 \pm 0.2$ & 13.9 \\
 & $f_2$ & $5.7941 \pm 0.0004$ & $5.0 \pm 0.2$ & 6.9 \\
\hline
\end{tabular} 
\tablefoot{\# is the WEBDA identification of the stars. The error estimates are formal and should be used with caution.}
\end{center}
\end{table}

\subsection{A $\beta$ Cephei suspect and an apparently constant star 
in the $\beta$ Cephei domain}

The light curves of NGC 637 6 (Fig.\ 7) raise the suspicion that it is a 
low amplitude $\beta$~Cephei star. In almost all nights, there is some 
variability on the time scale expected for pulsation, and the star is 
well separated from others in the field, implying no blending problems 
in the photometry. However, the Fourier analysis is inconclusive as the 
light curves are dominated by nightly variations of the mean stellar 
magnitude, most likely of instrumental origin\footnote{Nightly zeropoint 
variations at the same level are also present in the data for the other 
bright stars, but are small compared to their intrinsic variability.}. 
Still, the rms scatter per single point in this light curve is only 
2.6~mmag. Individual $\beta$~Cephei pulsation modes, if present, would 
not exceed a level of 1.5 mmag in amplitude.

\begin{figure}
\includegraphics[width=88mm,viewport=-0 00 310 165]{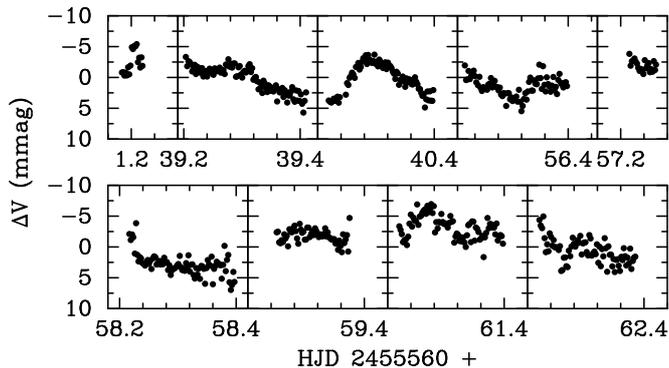}
\caption{Light curves of the $\beta$~Cephei suspect NGC~637~6.}
\end{figure}

Finally, NGC~637~137 did not show any sign of intrinsic variability. The 
star is a close double ($\approx10\arcsec$ separation) with NGC~637~138 
in the very centre of the cluster. Our images under best seeing suggest 
another faint star in between the two. In any case, the photometry of 
NGC~637~137 became poor under poor seeing conditions, but even taking 
only the best light curves (about 2/3 of the data) shows no trace of 
variability within a limit of 3 mmag in total or 1 mmag in the 
$\beta$~Cephei frequency domain.

\subsection{Other (claimed) variable and interesting stars in the 
field}

There are some other objects in the field of NGC 637 that deserve 
special attention. These are briefly discussed here, together with newly 
detected variables from the present study.



NGC~637~13 is variable in the present data, with a main frequency of 
2.46 \cd and an amplitude of 17~mmag.
Star WEBDA 18 was suggested to be variable (Huestamendia, del Rio \& 
Mermilliod \cite{H91}). The present measurements show no evidence for 
variability within a limit of 5 mmag, or 2 mmag for periods shorter than 
2.5~hr.

NGC~637~27 is a Be star (Yadav et al. \cite{Y08}). As these stars often 
are variable, we consequently examined our data. However, no variability 
within a limit of 5 mmag, or within 1.4 mmag for periods shorter than 
2.5~h, was found.

WEBDA 33 in NGC 637 is also variable, with a dominant frequency of 0.703 
\cd. The light curve is asymmetric, as also implied by the presence of a 
first harmonic from combining Fourier and residualgram methods, and 
reaches a total light range of 0.053 mag.

NGC~637~68 showed an egress from an eclipse in one night of measurement 
(Fig.\ 8). Finally, two new $\delta$ Scuti stars were discovered, 
GSC~04039$-$00930 with a dominant frequency of 14.36 \cd (modulo daily 
aliasing) and an amplitude of about 2.5\,mmag, and GSC~04040$-$01606 with 
apparent multiple frequencies between 6 and 8 \cd with amplitudes around 
12\,mmag.

\begin{figure}
\includegraphics[width=88mm,viewport=-0 00 310 090]{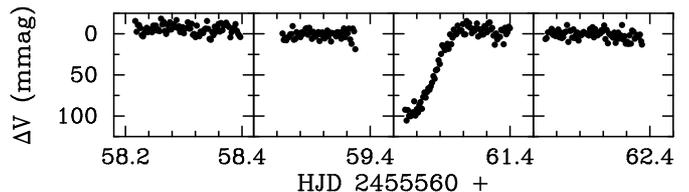}
\caption{Some light curves of the eclipsing binary NGC~637~68.}
\end{figure}

The positions of the new variables that have WEBDA identifications in a 
$V$ vs.\ $B-V$ colour magnitude diagram, constructed with WEBDA, are 
consistent with cluster membership. On the other hand, the two 
$\delta$~Scuti stars have no WEBDA designations and no accurate 
$BV$ photometry. Their mean $V$ magnitudes are $\approx 12.7$ and 
$\approx 13.9$ on our frames, respectively. In the following section, we 
will estimate a distance modulus of $V-M_V=13.6$ to NGC~637, implying 
$M_V=-0.9$ and 0.3, respectively, if the two pulsators were cluster 
members. The period-luminosity-colour relation for $\delta$~Scuti stars 
(Breger \cite{B79}) provides $M_V=1.7\pm0.7$ and $0.7\pm0.7$, 
respectively. The error estimates are dominated by the unknown colours 
of the variables and represent the width of the instability strip. We 
conclude that GSC~04039$-$00930 is a foreground star. GSC~04040$-$01606 may 
be a cluster member, although it is some $9\arcmin$ off the centre. It 
would then be a pre-main sequence pulsator because of its long periods 
and the young age of the cluster.

\section{Discussion}

The Str\"omgren $uvby\beta$ photometry by Handler (\cite{H11}) can be 
used to derive $T_{\mathrm{eff}}$ and $M_V$, thus log~$L$ of the 
individual stars, but also their reddenings and distance moduli. To this 
end, the routines by Napiwotzki et al.\ (\cite{NSW93}) were employed. 
Bolometric corrections by Flower (\cite{F96}) and a bolometric magnitude 
of $M_{\rm bol}=4.74$ for the Sun (Livingston \cite{L00}) were used to 
obtain stellar luminosities. The results of this procedure are listed in 
Table~4 and were also used for constructing Fig.\ 1.

\begin{table}
\begin{center}
\caption{Parameters of the seven brightest members of NGC~637 derived 
from Str\"omgren photometry.}
\begin{tabular}{ccccccc} 
\hline
\# & $T_{\mathrm{eff}}$ & log $g$ & $E(b-y)$ & $M_V$ & log $L$ & $V-M_V$ \\
 & (kK) & (dex) & (mag) & (mag) & ($L_{\sun}$) & (mag) \\
\hline
1 & 25.5 & 3.6 & 0.470 & $-$3.51 & 4.29 & 13.49\\
3 & 22.6 & 4.2 & 0.470 & $-$2.39 & 3.72 & 12.97\\
4 & 23.6 & 3.9 & 0.499 & $-$2.93 & 3.98 & 13.71\\
6 & 25.9 & 3.9 & 0.487 & $-$3.29 & 4.21 & 13.64\\
7 & 24.3 & 4.0 & 0.469 & $-$2.91 & 4.00 & 13.58\\
137 & 22.4 & 4.4 & 0.451 & $-$2.13 & 3.61 & 12.92\\
138 & 24.9 & 3.9 & 0.439 & $-$3.06 & 4.08 & 13.22\\
\hline
\end{tabular} 
\tablefoot{\# is the WEBDA identification of the stars. The error 
on $T_{\mathrm{eff}}$ is about $\pm 4$\%, and 0.2 dex in log g.}
\end{center}
\end{table}

The interstellar reddening of these stars is virtually the same, 
suggesting they are all at the same distance and are therefore cluster 
members. The mean value is $E(b-y)=0.469\pm0.008$\,mag, translating into 
$E(B-V)=0.65\pm0.01$\,mag. This is in excellent agreement with the value 
by Yadav et al.\ (\cite{Y08}), but less consistent with Pietrukowicz et 
al.\ (\cite{P06}).

Looking at the apparent distance moduli $V-M_V$, there are considerable 
differences of up to 0.79~mag between individual stars. However, it must be 
kept in mind that the absolute magnitude calibration of Str\"omgren 
photometry treats each star as if it was single. The spread in $V-M_V$ 
is very close to the absolute magnitude difference of a single star and 
a binary with equal components. It is therefore possible that stars 
WEBDA 3, 137 and 138 are members of binary systems. In fact, WEBDA 3 
probably shows ellipsoidal variations due to a secondary component. In 
any case, determining a cluster distance from the four stars with 
consistent $V-M_V$ (with a mean value of 13.6 mag) results in $d 
\approx 2.1$\,kpc, in reasonable agreement with literature results given 
our crude approach.


Turning to the $\beta$~Cephei stars, NGC 637 4 is among the five highest 
amplitude pulsators of this type ever discovered (cf.\ Pigulski \& 
Pojma{\'n}ski \cite{PP08}). Such high amplitude $\beta$~Cephei stars 
often show a ``stillstand" phenomenon in the rising branch of their 
light curves. This is not the case for NGC 637 4. For this star, a first 
harmonic was detected, but it modifies the light curve shape from a 
sinusoid towards a higher and sharper maximum and a flatter and 
shallower minimum, such as observed in pulsating white dwarf stars 
(e.g., Montgomery \cite{M05}).

From an asteroseismic point of view, the large amplitudes of the two 
dominant modes of NGC 637 4 easily allow mode identification from 
multicolour photometry, once resolved from each other. The strongest 
modes in the three new $\beta$~Cephei pulsators WEBDA 1, 7, and 138 
reported here may also be identified with a rather modest data set.

Considering the seven brightest stars in this cluster and their light 
variability as an ensemble, four are now known as $\beta$~Cephei stars, 
and one is a suspect. There are currently too few observational 
constraints to pinpoint the distinction between pulsators and 
nonpulsators. However, it is interesting to note that the two apparently 
pulsationally stable stars may have binary companions, as implied by the 
difference between their apparent and absolute magnitudes from 
Str\"omgren photometry. They would therefore be less massive than 
the pulsators. More evidence in this direction would be provided by 
confirmation of ellipsodial variability of WEBDA 3.

\section{Summary and conclusions}

Thirty-one hours of time resolved $V$ filter CCD photometry of the open 
cluster NGC~637 resulted in the confirmation of two previously known 
variable cluster members, one $\beta$~Cephei pulsator and one probable 
ellipsoidal variable. Three new $\beta$~Cephei stars were discovered in 
the cluster and one more star is suspected to show low amplitude 
oscillations of the same kind. Other variable stars in the field are 
also reported. The present study shows that precise relative photometry 
can be obtained with a small telescope at an urban site.

NGC~637 is now among the six open clusters with most $\beta$~Cephei 
stars known (together with NGC 884, NGC 3293, NGC 4755, NGC 6231 and NGC 
6910, Pigulski \cite{P04}, Stankov \& Handler \cite{SH05}, Saesen et al. 
\cite{SP10}). The presence of at least four pulsators within a small 
portion of the sky (the seven brightest stars are located within 
$3\arcmin \times$\,$4\arcmin$), the large amplitudes of the dominant 
modes and the apparent presence of nonpulsating stars within the 
$\beta$~Cephei domain make NGC~637 an attractive target for ensemble 
asteroseismology. Time-resolved photometric observations in several 
filters, including an ultraviolet band, for pulsational mode detection 
and identification, are easily possible. Spectroscopic observations of 
the seven brightest cluster members are desirable for several purposes: 
to determine the rotational velocities, to look for binary companions 
and/or radial velocity variations, to derive more accurate $T_{\mathrm{eff}}$ 
and log~$g$ values, and perhaps even to determine abundances of chemical 
elements.

\begin{acknowledgements}
This research is supported by the Austrian Fonds zur F\"orderung der 
wissenschaftlichen Forschung under grant P20526-N16. This research has 
made use of the WEBDA database, operated at the Institute for Astronomy 
of the University of Vienna. We are grateful for the constructive 
comments of the referee.
\end{acknowledgements}

\end{document}